\documentclass[aip,
 amsmath,amssymb,
 reprint,%
]{revtex4-1}

\usepackage{graphicx}
\usepackage{dcolumn}
\usepackage{bm}

\usepackage{mathptmx}
\usepackage{etoolbox}

\makeatletter
\def\@email#1#2{%
 \endgroup
 \patchcmd{\titleblock@produce}
  {\frontmatter@RRAPformat}
  {\frontmatter@RRAPformat{\produce@RRAP{*#1\href{mailto:#2}{#2}}}\frontmatter@RRAPformat}
  {}{}
}%
\makeatother

\usepackage{amsmath}
\usepackage{amssymb}
\usepackage{graphicx}
\usepackage{dcolumn}
\usepackage{bm}
\usepackage{multirow}
\usepackage{hyperref}
\usepackage{color}
\usepackage{xcolor}
\usepackage{appendix}

\newcommand{\rev}[1]{\textcolor{black}{ #1 }}
\newenvironment{rev_par}
  {\begingroup\color{black}}
  {\endgroup}

\begin{document}

\title{Structural and Compositional Complexities of Hierarchical Self-Assembly: a Hypergraph Approach }

\author{Alexei V. Tkachenko}
\email{oleksiyt@bnl.gov}
\affiliation{Center for Functional Nanomaterials, Brookhaven National Laboratory, Upton, NY 11973, USA}

\begin{abstract}
Programmable self-assembly enables the construction of complex molecular,
supramolecular, and crystalline architectures from well-designed building
blocks. We introduce a hypergraph-based formalism, \textit{Blocks \& Bonds}
(B\&B), that generalizes classical chemical graph theory by incorporating
directed and multicolored interactions, internal symmetries, and hierarchical
organization. Within this framework, we develop the \textit{Structure Code}
(SC), a compact and versatile  language for describing self-assembled
architectures.
\rev{
We define a Kolmogorov-style \textit{Structural Complexity} as the total
information content of SC, obtained through its tokenization and
Shannon information assignment. Complementing this encoding-based measure, we
introduce a much simpler quantity, the \textit{Compositional Complexity}, which
depends only on the number and cumulative usage of block and bond types in the
construction set.
A central result of this work is a strong empirical correlation between the
token-based Structural Complexity and the Compositional Complexity across all
examined systems. Owing to this agreement, the Compositional Complexity
emerges as the most practical and broadly applicable measure: it is easy to
compute, requires no explicit encoding, and yet closely tracks the actual
information content of structurally diverse architectures.}
Applications to molecular systems (ethylene glycol, glucose), DNA-origami
lattices, and crystalline assemblies show that B\&B hypergraphs provide a
unified, scalable, and information-efficient representation of structural
organization, naturally capturing symmetry, modularity, and stereochemistry.
This framework establishes a quantitative foundation for complexity-aware
classification and inverse design of programmable matter.

\end{abstract}

\maketitle

\section{Introduction}

Programmable self-assembly has emerged as a powerful paradigm in nanoscience and materials engineering. Enabled by advances in DNA nanotechnology, colloidal design, and and bio-mimetic chemistry, researchers can now engineer building blocks with highly specific interactions that drive the autonomous formation of target structures \cite{ke2012three, halverson2013dna,PhysRevLett.112.238103,jacobs2016self,Gang_Tkachenko_2016,ACS2024,videbaek2024economical,Jason2025}. This capability has transformed the field from descriptive studies of natural assemblies to predictive and design-oriented frameworks for artificial matter. Recent successes include addressable DNA origami lattices, patchy colloids with programmable valences, and hierarchically organized supramolecular materials.

\rev{
These developments expose a fundamental question: \emph{how should one quantify the complexity of a self-assembled structure?} As structures become increasingly intricate, multicomponent, and hierarchical, traditional descriptors—bond graphs, coordination environments, or space-group symmetries—are no longer sufficient to capture their true information content. A meaningful complexity measure must simultaneously reflect the diversity of building blocks, the specificities of their interactions, and the compressibility of the structure arising from symmetries, repetitions, and hierarchical reuse of subunits.
}

These advances highlight the long-standing challenge of defining the very concept of complexity: as assemblies become more intricate, how can one quantify and compare their structural information content? Kolmogorov defined the complexity of an object, such as a text, as the length of a computer program that can generate that object as an output \cite{kolmogorov1965,kolmogrov1968}. In practice, exact Kolmogorov complexity is uncomputable, but useful proxies can be constructed through compression algorithms, such as the Lempel–Ziv–Welch (LZW) family \cite{LZ78,LZW}, which generate a dictionary of repeating motifs and underlie common compression formats including \text{gzip}, \text{zip}, \text{tiff}, \text{gif}, and \text{png}.

In the context of chemistry and materials science, a variety of complexity measures have been proposed, but none have been universally adopted \cite{Mao2024}. A plausible candidate is the so-called molecular information content originating from Rashevsky’s \cite{rashevsky1955life} and Mowshowitz’s \cite{mowshowitz1968entropy} work on graph entropies, later refined by Bonchev \cite{bonchev1976symmetry}, Bertz \cite{BERTZ1983849,Bertz1981}, and others \cite{dehmer2011history,dehmer2012revisited,krivovichev2012topological,krivovichev2014inorganic,Sabirov2021}. 

\rev{
However, these classical measures were designed for simple molecular graphs and therefore fail to capture essential features of modern programmable assemblies: multi-terminal building blocks, patch specificity, internal and emergent symmetries, and hierarchical modularity. A modern complexity measure must generalize beyond pairwise graphs and account for the nested, multi-layered structure typical of DNA constructs, patchy colloids, addressable lattices, and supramolecular architectures.
}

In this work, we propose a framework for encoding and quantifying the complexity of programmable self-assembled structures using \textit{hypergraphs} \cite{hypergraphs}. Extending conventional chemical graph theory, our \textit{Blocks \& Bonds (B\&B) hypergraphs} represent composite units and multiple interaction types while naturally capturing nested and hierarchical organization. We further define structural complexity as the minimal information required to encode such hypergraphs, drawing explicit connections to Kolmogorov and information-theoretical perspectives. This approach allows a classification of complexity across molecules, supramolecular assemblies, and synthetic hierarchical materials.

\section{From graphs and hypergraphs}

\rev{
Graph theory provides a powerful language for describing complex, network-like organizations across many scientific fields. In chemistry and materials science, graph-based models have long been used to represent the connectivity and structural complexity of molecules and crystalline solids. In a chemical graph, atoms correspond to vertices and bonds correspond to edges. When a structure is sufficiently symmetric, its graph admits automorphisms, i.e., mappings of the graph onto itself that preserve all connectivity relations. Such automorphisms naturally group atoms into equivalence classes: sets of vertices that can be permuted among one another without changing the graph. These classes correspond to chemically equivalent atomic sites.
}

Suppose a molecule contains (M) atoms that fall into (K) equivalence classes, with $(m_1,\dots,m_K)$ atoms in each class. Rashevsky (1955) \cite{rashevsky1955life} proposed the following expression as a measure of a chemical graph’s “information content” or complexity:
\begin{equation}
H = -\frac{1}{M}\sum_{i=1}^K m_i \log_2\left(\frac{m_i}{M}\right).
\label{eq:I}
\end{equation}
The form of this expression closely resembles Shannon’s entropy, and this analogy motivated its early adoption. Over the decades, it has been widely used and generalized in the work of Mowshowitz \cite{mowshowitz1968entropy}, Bonchev \cite{bonchev1976symmetry}, Bertz \cite{Bertz1981,BERTZ1983849}, and many others \cite{dehmer2011history,dehmer2012revisited,krivovichev2012topological,krivovichev2014inorganic,Sabirov2021} as a general indicator of molecular or topological complexity.
\rev{However, the actual meaning of Rashevsky’s information content has remained ambiguous. In essence, Eq. (\ref{eq:I})  describes how much information, on average, is required to specify which equivalence class (i.e. a chemically equivalent site) a randomly selected atom belongs to.
What it does not describe is the amount of information required to encode the molecular graph itself. To our knowledge, no rigorous connection has been established between Rashevsky’s formula and the minimal information needed to represent a graph uniquely. This limitation was noted by Bertz  \cite{Bertz1981,BERTZ1983849}, who ultimately abandoned the information-theoretic interpretation and proposed an alternative measure that incorporates additional topological features of the graph. The resulting “Bertz complexity” remains widely used, despite the lack of its clear interpretation. }

While chemical graphs provide a useful abstraction framework, they face several limitations when applied to more complex architectures such as supramolecular complexes or programmable self-assembly. In particular, standard graphs treat atoms as indivisible vertices, but ignore geometrical aspects of their binding, as well as the diversity of bond types often employed in programmable self-assembly.  
In addition, real self-assembled systems often involve composite units (functional groups, colloids, DNA tiles) with a non-trivial internal organization. Graph-based complexity measures capture global automorphisms but largely ignore such nested or recursive organizational levels.

To address these challenges, we generalize the representation of complex structures by employing {\it hypergraphs} \cite{hypergraphs} rather than standard graphs. A hypergraph is defined as a pair $(V,E)$, where $V = \{v_1,\dots,v_N\}$ is the set of vertices and $E$ is a collection of subsets of $V$, called {\it hyperedges}. Unlike ordinary edges, hyperedges may connect any number of vertices simultaneously.

As a generalization of molecular graphs, we introduce {\it Blocks \& Bonds (B\&B) Hypergraphs}. In the B\&B framework, two types of hyperedges are distinguished:
\begin{itemize}
\item {\it Blocks}: hyperedges that group together $k$ vertices. Each block is endowed with a group $G\subseteq S_k$, where $S_k$ is the symmetric group on $k$ elements, which defines equivalence among vertex permutations.
\item {\it Bonds}: pairwise connections $e_{ij}(v_i,v_j)$, subject to the restriction that each vertex may participate in at most one such bond.  
\end{itemize}

\begin{figure}
    \centering
    \includegraphics[width=\linewidth]{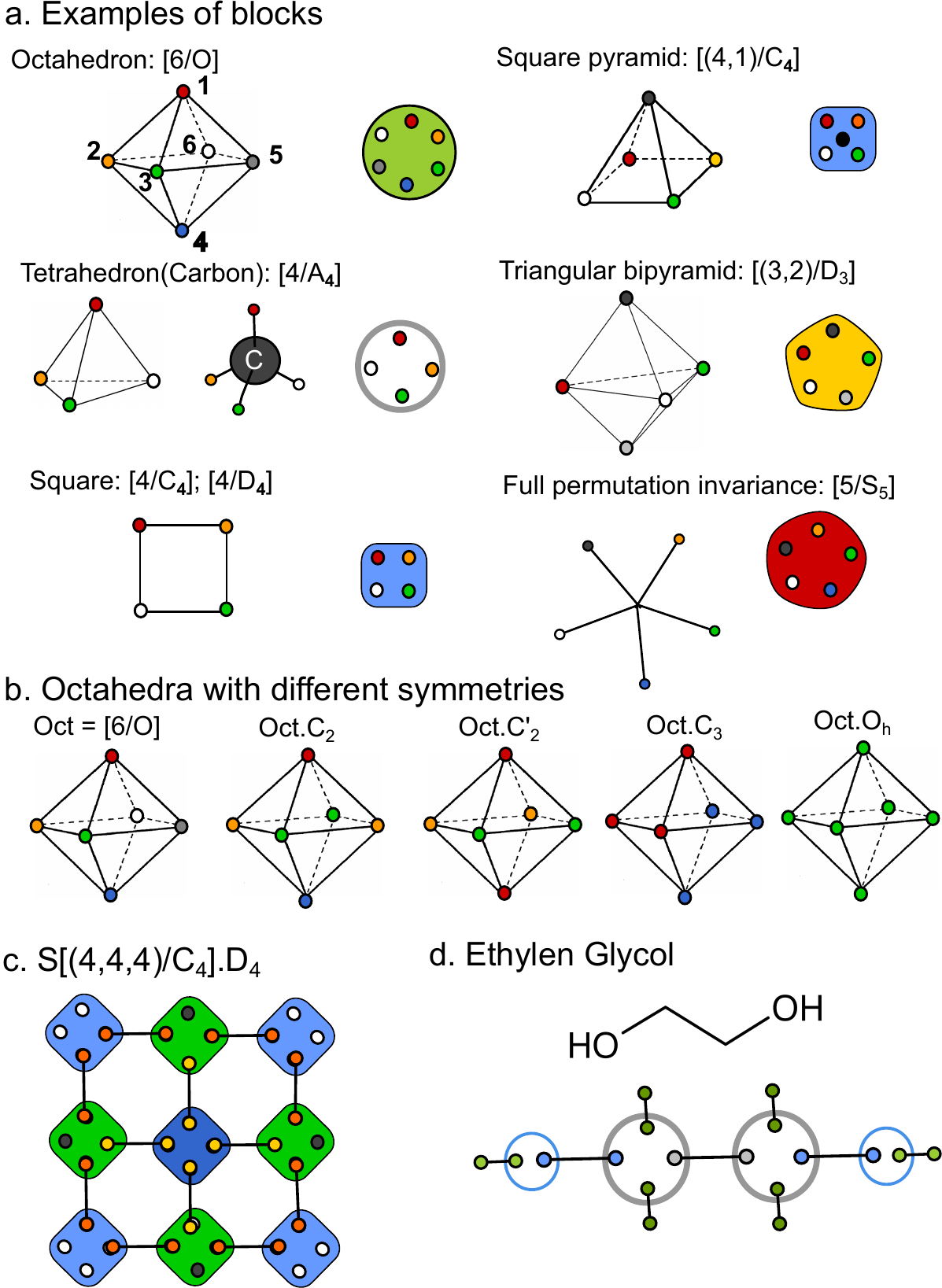}   
    \caption{Examples of Blocks \& Bonds (B\&B) hypergraph representations.  
    (a) Polyhedral blocks such as octahedron $[6/O]$, tetrahedron $[4/A_4]$, and square $[4/C_4]$.  
    (b) Variants of octahedral blocks with different labeling symmetries.  
    (c) Composite hypergraph structure constructed from square blocks.  
    (d) Molecular hypergraph representation of ethylene glycol, $(CH_2OH)_2$.}
    \label{fig:Fig1}
\end{figure}

Vertices of a block not involved in internal bonds are called {\it terminals}. Each block is characterized by its {\it valence} $k$ (the number of terminals) and by the equivalence group $G$, which is a subgroup of $S_k$, the full permutation group on $k$ elements:  $G \subseteq S_k$. The group $G$ defines which permutations of terminals are considered equivalent. Such a block is denoted
$B = [k/G]$.   Several familiar polyhedral structures can be expressed in this way, as shown in Figure~\ref {fig:Fig1}a.:
\begin{itemize}
\item  A block $[6/O]$ represents the six vertices of an octahedron. The $6!$ possible permutations are partitioned into 30 equivalence classes of 24 permutations each, related by octahedral rotations, e.g. $(x,y,z,x^*,y^*,z^*) \sim (x,z,y*,x^*,z^*,y) \sim (y,z,x,y^*,z^*,x^*) \sim \dots$.
\item  A tetrahedron corresponds to $[4/A_4]$, since any even permutation of four vertices (group $A_4$) corresponds to a 3D rotation.
\item  A square can be represented as $[4/C_4]$ (cyclic rotations only) or $[4/D_4]$ (including reflections).
\end {itemize}
In some cases, group actions split the vertices into independent subsets, or orbits, with different multiplicities. For such cases, we will use notation
\begin{equation}
B = [(k_1,\dots,k_l)/G]    
\end{equation}
where the $k_i$ denote orbit sizes. For instance, as shown in Figure  ~\ref{fig:Fig1}a., a square pyramid has four basal vertices permuted by $C_4$ and one apex fixed by the group, giving $[(4,1)/C_4]$. A triangular bipyramid corresponds to $[(3,2)/D_3]$, with three equatorial and two apical vertices. 

If no symmetry is imposed, the trivial group is assumed, and all permutations are distinct. A block $[k/S_k]$ with full permutation symmetry is equivalent to a vertex of degree $k$ in the standard conventional graph theory. 
In the context of m, molecular hypergraphs, atoms of low valence correspond to symmetric blocks, e.g., oxygen $[2/C_2]$ or nitrogen $[3/S_3]$. For higher valence, only certain permutations are realizable by spatial rotations; e.g., for  $sp^3$ hybridized carbon,   the tetrahedral arrangement of valence electrons corresponds to block $[4/A_4]$.  

\begin{rev_par}
It is essential to distinguish between a block’s \emph{equivalence group} and its
\emph{symmetry group}, as these two concepts play different roles in the
B\&B formalism.  

\textbf{The equivalence group} \(G\subseteq S_k\) is a formal property of the
\emph{block definition}.  It specifies which permutations of the $k$ terminals
are regarded as representing the \emph{same block type}.  For example, the
octahedral block $[6/O]$ identifies all vertex labelings that differ only by
rigid rotations of an octahedron.  Thus $G$ determines how terminals may be
permuted without changing the identity of the block.

\textbf{The symmetry group} $\mathrm{Sym}$, in contrast, depends on the
\emph{actual labels} assigned to the terminals when the block is instantiated
inside a structure.  It consists of permutations that preserve both the
geometry of the block and the specific labeling.  
These notions are illustrated by octahedral examples in
Fig.~\ref{fig:Fig1}b. An octahedron with six distinct terminal labels
$(x,y,z,x^*,y^*,z^*)$ has equivalence group $O$, but its symmetry group is 
trivial, since no nontrivial permutation preserves the labeling.  If all six
labels are identical, the symmetry becomes the full $O_h$, which includes
reflections absent from $O$.  A mixed labeling such as $(a,b,b,b,b,a)$ yields
a symmetry group $D_4$, which is a subgroup of $O_h$ but not of $O$.

To capture both aspects explicitly, we denote a block with equivalence group
$G$ and symmetry group $\mathrm{Sym}$ as $B = [k/G].\mathrm{Sym}$.  This
notation makes clear that $G$ constrains the \emph{permissible terminal
permutations}, while $\mathrm{Sym}$ encodes the \emph{actual symmetry} of the
labeled instance used in a structure.

\end{rev_par}

\section{Structure code}
While B\&B hypergraphs provide a versatile structural representation, they do not by
themselves specify a unique encoding of a structure. To quantify information content, we 
require a concrete and unambiguous textual representation. Below, we present  Structure Code, 
a compact language 
whose syntactic structure mirrors the hierarchical organization of B\&B 
hypergraphs, enabling Kolmogorov-style analysis of Complexity.

A block will be called fundamental if all of its vertices are terminals, i.e., it contains no internal bonds. A structure is specified by listing its fundamental blocks together with their bonds. Naturally, highly symmetric structures can be encoded with fewer block and bond types.   In general, a composite block can be expressed in the following Block--Vertex--Multiplicity--Symmetry ({\it BVMS})  format:
\begin{align}
\nonumber B=[B1(v_1,..,v_l)^{m1},(v_{l+1},..,v_j)^{m2}...\\
...~|~ B2. ...~|~ B3. ....].Sym
\end{align}
Here, $B1,B2,B3,..$ are predefined sub-blocks, fundamental or composite;  $v_i$ are vertex types and $m1,m2, ..$ are multiplicities of the specific blocks, and $Sym$ is the symmetry group of the composite block $B$. There is a natural correspondence between the vertex types used in the structure code and bond "colors" in the context of programmable self-assembly. The latter are typically encoded by mutually complementary DNA chains attached at specific locations of DNA origami building blocks.  This type of description can be scaled up to encode an arbitrary B$\&$B hypergraph with a Structure Code (SC). \rev{Its brief description is presented at the end of this section, and full specifications are given in Appendix \ref{app:structurecode}. Below, we illustrate its key elements on several examples.}

 Our first example, consider the composite structure shown in Fig.~\ref{fig:Fig1}c.  It has the following {\it SC}:
\begin{align}
\nonumber &SQ = [4/C_4];~ SQ_4=SQ.D_4\\
 &SQ_2=SQ.C_2(a,*,a,*);~ SQ'_2=SQ.C'_2(a,a,b,b) \label{eq:square}\\
\nonumber &S  = [SQ_2(1^*,2^*,*)^4 ~|~ SQ'_2(1,*)^4 ~|~ SQ_4(2)].D_4
\end{align}
Here $1,2,1^*,2^*$ denote vertex types, with $(v,v^*)$ designating bond-forming pairs, and $*$ marking terminals with no designated type. We used block symmetries for a compact listing of vertex types.  In this construction, all fundamental blocks are squares with equivalence group $C_4$. The assembled structure possesses global $D_4$ symmetry, while individual blocks differ: the central square $SQ.D_4$ is fully symmetric, four side blocks exhibit mirror symmetry $C_2$, and four corner blocks also have mirror symmetry $C'_2$, but with respect to a different axis. Overall block $S$ is of type $[(4,4,4)/C_4].D_4$, which means that 12 terminals can be split onto 3 orbits of the equivalence group $C_4$. If no symmetry was specified, that would allow all the terminals to be distinct, subject to the equivalence relationship $(1,2,3,4,5,6,7,8,9,10,11,12)\sim(4,5,6,7,8,9,10,11,12,1,2,3)$. However, due to symmetry $D_4$, there are only two terminal types: $D4(a,b)=(a,b,a,a,b,a,a,b,a,a,b,a)$.

Another example is a molecular hypergraph of Ethylene glycol, $(CH_2OH)_2$ shown in Figure ~\ref{fig:Fig1}d:  
\begin{align}
     & \label{eq:EG}  C=[4/A_4].C2;~~ O=[2/C_2];~~ H=[1]~~ \\
   & \nonumber EG=[C(0,2,3)^2 ~|~ O(2^*,1)^2 ~|~ H(0^*)^4,(1^*)^2].C_2  
\end{align}
Here and below, the absence of a complementary vertex to type $3$ indicates that this type is self-complementary, i.e. the bond is formed between two vertices of the same type.

\begin{figure}
    \centering
    \includegraphics[width=\linewidth]{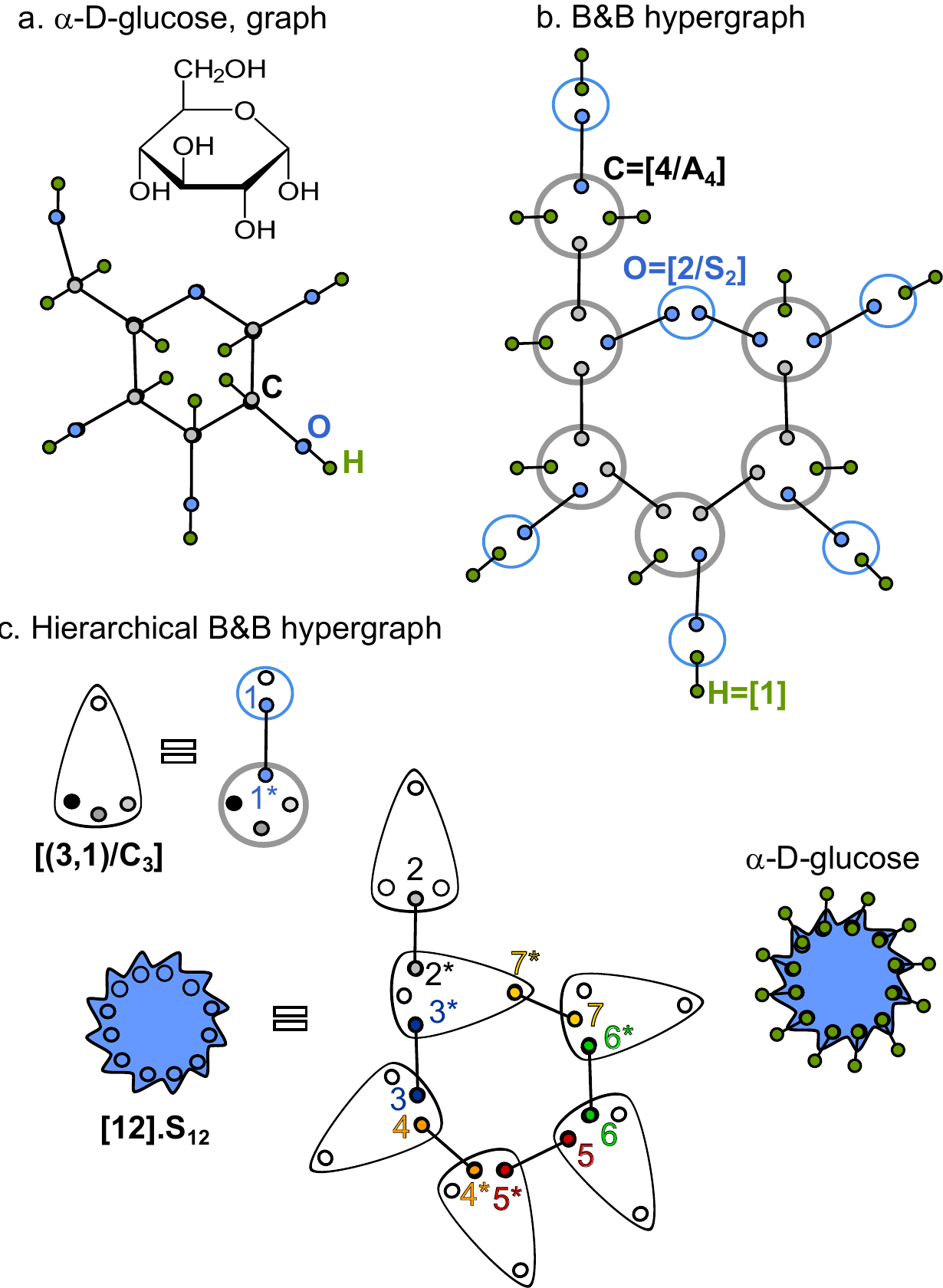}   
    \caption{Hypergraph representations of $\alpha$-D-glucose.  
    (a) Conventional graph representation.  
    (b) B\&B hypergraph.  
    (c) Hierarchical encoding using composite $C$–$O$ blocks, enabling compact representation of repeated motifs.}
    \label{fig:Fig2}
\end{figure}

The B\&B hypergraphs provide a natural framework for describing structures with hierarchical organization. That is because any subgraph that can be isolated from the rest of the hypergraph by cutting its outside bonds is itself a block. This allows for a more compact encoding even when no true symmetries are present. As an example, consider a somewhat more complicated molecule than before, $\alpha$-D-glucose, shown in Figure \ref{fig:Fig2}.  Panels a and b represent this molecule as a graph and a B\&B hypergraph, respectively. An important advantage of the B\&B hypergraph is that it allows us to distinguish specific stereoisomers (e.g. $\alpha-D$ vs $\beta-D$ versions of glucose), a feature that is captured by a simple graph representation.  Neither the molecule itself no its hypergraph has any global symmetry.  The graph has a single $C_2$ automorphism, which corresponds to swapping the two hydrogen atoms that belong to the side chain. This implies a relatively high complexity according to the Rashevsky formula,  as well as a long and tedious SC. However, one can notice structural repeats within the glucose molecule. For instance, one can define a composite block $C-O$, as shown in Figure  \ref{fig:Fig2}c. The type of this block, $[(3,1)/C_3]$ is dictated by its constituents: once Oxygen is bound to $C$, the remaining 3 terminals of $C$ preserve the cyclic equivalence group $C_3$. Now, $\alpha$-D-glucose can be composed of these blocks, and its structure can be encoded as:
\begin{align}
\label{glu}
&C=[4/A_4];~~ O=[2/C_2];~~ H=[1]\\
\nonumber &CO=[C(*,*,*,1^*)~|~ O(1,*)  ]\\
\nonumber &G= [CO(2,*,*,*),(2^*,*,3^*,7^*),(*,3,4,*),\\
\nonumber  &(*,4^*,5^*,*),(5,6,*,*),(6^*,7,*,*)  ].{S_{12}} \\
\nonumber &Glc_\alpha= [G(0)~| ~H(0^*)^{12}]
\end{align}

Note that once we encoded the ring structure out of 6 $CO$ blocks, all 12 of its terminals have been assigned the same type $0^*$, by  applying $S_{12}$ permutation group. This is not, of course, the symmetry group of the structure itself, but a compact way of encoding repeated vertex types. In fact, each type of monovalent (terminal) group, such as Hydrogen $H$, only requires a single bond type to encode its positioning anywhere in a hypergraph. 

\begin{rev_par}

As another demonstration of the power of hierarchical design, consider the classical C$_{60}$ fullerene structure, i.e., the truncated icosahedron. Since this is a known allotrope of carbon, it may be tempting to use a minimalistic code featuring a single trivalent block type, with all blocks connected by the same bond type:
$
C = [3/S_3], ~ C60 = [ C(1)^{60} ]. I_h 
$. 
This code would indeed unambiguously select the C$_{60}$ cluster from among multiple carbon nanostructures, since it is the only one containing $60$ atoms and possessing icosahedral symmetry $I_h$. However, in a broader context, the code is ambiguous: there exists another geometrically plausible structure with the same symmetry, composition, and connectivity: a truncated dodecahedron (see Figure  \ref{fig:Fig3}). Thus, a more complex structural code is needed to uniquely identify C$*{60}$. Once again, we employ a hierarchical construction. Starting with a lower-symmetry trivalent block “Y”, we define a pentagonal composite block $P$, as shown in Figure  \ref{fig:Fig3}a, and then assemble the final cluster from twelve copies of it:
\begin{align}
 &Y= [(2,1)/C_2]. C_2; ~P= [ Y(1,*)^{5} ]. C_5 \nonumber\\
 &C60 = [ P(2)^{12} ]. I_h \label{eq:C60}
\end{align}
A similar hierarchical code based on triangular motifs, $T=[ Y(1,*)^{3} ]. C_3$,  is possible for the truncated dodecahedron, as shown in  Figure  \ref{fig:Fig3}a.
\begin{figure}
    \centering
    \includegraphics[width=\linewidth]{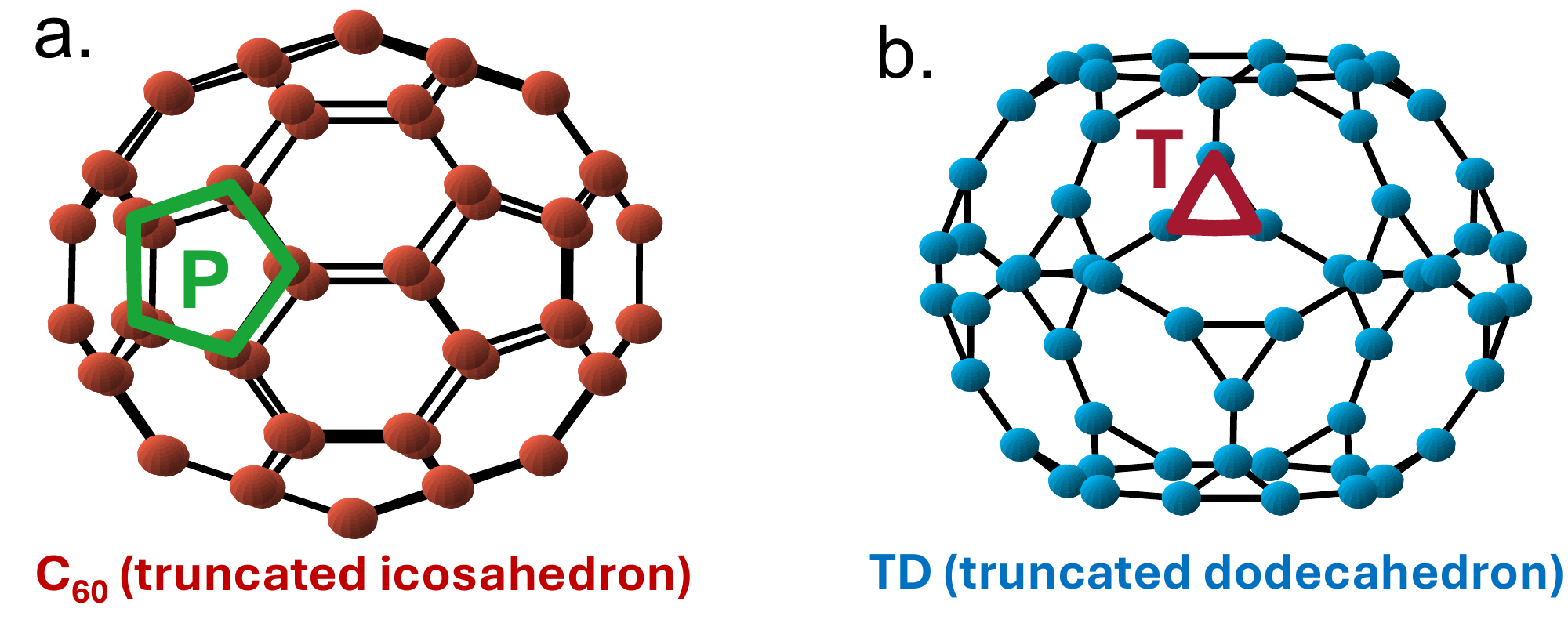}
    \caption{\rev{Two structures, both containing 60 trivalent blocks, and possessing icosahedral symmetry: (a)  C$_{60}$ fullerene, and (b) Truncated dodecahedron (TD). 
  }
    }
    \label{fig:Fig3}
\end{figure}

\end{rev_par}

The SC can also be employed for infinite crystals. As an example, consider designs for programmable self-assembly generated by MOSES  algorithm, described in Ref. \cite{Jason2025}. It was developed specifically for the self-assembly platform based on multi-chromatic octahedral DNA origami(i.e. $Oct=[6/O]$ in present notations). Each octahedral unit acts as a 3D pixel (voxel), allowing to the positioning of specific "cargo" at predetermined locations within a cubic lattice. For instance, below is the MOSES-generated design of Zinkblende ($ZnS$) structure, in which  two types of "cargo voxels" are positioned at the  sub-lattices within a cubic crystal that correspond to locations of $Zn$ and $S$, respectively:  
\begin{align}
   \nonumber &Oct=[6/O];~B,C1,C2=Oct.O_h\\
   \label{eq:ZB} &A=Oct.C_2(a,b,*,a,b,*)\\
   \nonumber &ZB=[A(5^*,2^*,3^*,1^*)^6,(5,2,6,4)^{6}~\\
   \nonumber &|~B(3),(4^*)~|~C1(1)~|~ C2(6^*)]^{\infty}.F\bar{4}3m  
\end{align}
 Here notations $[...]^\infty.Sym$ are used to indicate that the structure is a crystal, with crystallographic symmetry group $Sym$ specified.  Block multiplicities are given per (cubic) cell. Note that there are three distinct octahedral types and symmetries $O_h$: $B,C1, C2$. Two of them, $C1$ and $C2$, are cargo voxels that are arrange into the Zinkblende structure.  
 
 Another example from Ref. \cite{Jason2025} is the design for the cubic Laves phase, MgCu$_2$: 
\begin{align}
\nonumber &Oct=[6/O];~~B,C1,C2=Oct.O_h; ~~D=Oct.C_3;\\
\nonumber&A1=Oct.C_2C_2(a,b,*,a,b,*);~A2=Oct.C'_2(a,b,b,a,c,c); \\
\nonumber&LF=[Oct(11,13,14,12,3,2)^{12},(16,18,8,17,9,15)^{12}  \\ 
\nonumber&| ~A1(2^*,3^*,4^*,1^*)^6,(5,6,7,4)^6,(8^*,9^*,10^*,7^*)^6 \\
\label{eq:LF}&|~A2(6^*,15^*,14^*)^6, (5,18^*,13^*)^6~|~ D(17^*,11^*)^4 \\
\nonumber&|~B(10),(12^*)^2 ~|~ C1(1) ~|~ C2(16^*)^2]^\infty.Fd\bar{3}m
\end{align}

\begin{rev_par}
    
Below is a brief overview of the SC language. Its full specifications are presented in Appendix \ref{app:structurecode} 

\begin{enumerate}
    \item \textbf{Vertex types.}
    Terminals are assigned types $v$, $v^*$, or $*$.
    Complementary types $v$ and $v^*$ form a bond, while ``$*$'' denotes
    a non-binding terminal.

    \item \textbf{Block declarations.}
    A fundamental block of valence $k$ is declared as
    \[
        B = [k/G].\mathrm{Sym},
    \]
    where $G \subseteq S_k$ is an Equivalence group defining which
    permutations of terminals are considered identical  (multi-orbit blocks may be written as $[(k_1,\ldots,k_\ell)/G]$). 
    $\mathrm{Sym} $ is the Symmetry group of the block, that enforces equivalence of vertex types that belong to the same orbit of it. 

    \item \textbf{Composite blocks.}
    Larger units are encoded using the Block--Vertex--Multiplicity--Symmetry
    (BVMS) format:
    \[
        B = [\, B_1(v_1,\ldots)^{m_1} \;|\; B_2(\ldots)^{m_2}
        \;|\; \cdots \,].\mathrm{Sym}
    \]
    Each term specifies a sub-block $B_i$, the assignment of vertex types to
    its terminals, and its multiplicity $m_i$. Composite blocks may be nested
    hierarchically.

    \item \textbf{Bonds.}
    Bonds are implicit: each vertex type $v$ must pair exactly once with its
    complement $v^*$, unless explicitly redefined (e.g. $3 = 3^*$).

    \item \textbf{Crystalline structures.}
    Periodic assemblies are encoded per unit cell using
    \[
        [\, \cdots \,]^\infty.\mathrm{Sym}
    \]
    where $\mathrm{Sym}$ is a crystallographic space group in
    Hermann-Mauguin notation.
\end{enumerate}

\end{rev_par}

\section{Compositional Complexity}

Now that we have established a format for encoding a structure as a string, the length of this code provides a natural measure of its information content, and therefore of its  \textit{Kolmogorov Structural  Complexity} \cite{kolmogorov1965,kolmogrov1968}. \rev{ More precisely, this Complexity is defined as the \textit{length of the shortest SC that unambiguously specifies the structure in (3D) space}. Symmetry and hierarchical organization reduce this complexity by minimizing the number of distinct block and vertex types required to describe the structure, thereby shortening the code. Since any SC can, in principle, be compressed, its true information content is determined by the most efficient possible encoding \cite{coding}.
}

\begin{rev_par}

We now introduce a practical proxy for Kolmogorov-style Structural Complexity, which we refer to as the \emph{B$\&$B Complexity}. It estimates the minimal number of bits needed to encode all block and bond types appearing in a Sc. Based on  Shannon’s source coding theorem \cite{shannon}, this minimal information is given by
\begin{equation}
C_{B\&B} = 
\min_{\text{valid codes}}
\left[-\sum_{\alpha=1}^{\Omega}
I_\alpha \,
\log_2\!\left(\frac{I_\alpha}{N}\right)
\right],
\label{eq:C}
\end{equation}
where the minimization is taken over all SCs that uniquely specify the same physical structure.  
Here $I_\alpha$ is the \emph{Index} of   block or bond type $\alpha$:
\begin{itemize}
\item The \textit{index of a block} $B$,  is the number of occurrences of $B$ in the definitions of \emph{other} blocks.  
Introducing a block as a clone of an existing one (e.g.\ $A=B$) does not contribute to the index.  
\item The \textit{index of a bond type} $b$ is defined as the number of the corresponding vertex types appearing in the code. In all the codes presented above, this corresponds to   $I=2$ for bonds formed by different complementary vertices, $I=1$ for the self-complementary case, and $0$ for unbound void $*$ vertices. 
\end{itemize}
$\Omega$ in  Eq. (\ref{eq:C}) denotes the total number of block and bond types, and   
\begin{equation}
     N=\sum_{\alpha} I_\alpha
\end{equation}
is their cumulative index, i.e., the total number of occurrences in the code. 
\end{rev_par}

The apparent similarity between  Eq. (\ref{eq:C}), and the Rashevsky formula, Eq.  (\ref{eq:I}), is somewhat misleading.  While both are related to  Shannon's Information Entropy, the two results have very different origins and interpretations.  In particular, Rashevsky complexity is typically measured in bits/atom, and its cumulative value (obtained by multiplying  Eq. (\ref{eq:I}) by the number of atoms $M$) is extensive in the system size. In contrast, the B$\&$B Complexity is by definition already a cumulative value, and there is no compelling reason to calculate its per atom content:  $C_{B\&B}$ may well be finite even for an infinite system, such as a crystal. This is quite natural, since one does not require much information to encode an infinite repeat of a particular unit cell. In addition, the hypergraphs provide a much richer framework than regular graph theory. As a comparison, consider two representations of glucose shown in Fig. \ref{fig:Fig2} a-b, respectively. The graph contains 24 vertices, only 2 of which are in the same equivalence class. This gives, according to Eq.(\ref{eq:I}), information content of $H=\log_2 24 -1/12= 4.5$  bit/atom, or $108.0$ bits cumulatively. The hypergraph of the same structure, shown in Fig. \ref{fig:Fig2}b, contains 24 blocks and 48 bonds connecting 96 vertices, all of them distinct. This gives a naive result for B$\&$B complexity as high as $24\log_2 120=828.8$ bits. This dramatic increase is simply due to the higher complexity of the hypergraph representation: it has 1 block and 4 vertices per atom, as opposed to a single vertex in the graph. However, once modularity and hierarchical properties of the hypergraph are taken into account, as in  Eq.(\ref{glu}), the B$\&$B complexity gets reduced down to $76.2$ bits.             

\begin{rev_par}
We now introduce a compact book-keeping tool for recording the indices of a given structure, which we call the \textit{Index Monomial}. To construct it, we define the index multiplicity $\mu_I$ as the number of distinct vertex or block types that occur with multiplicity $I$.  The Index Monomial (IM) of an arbitrary structure is defined as
\begin{equation}
\mathcal{M} = \prod_{I=1}^{I_{\max}} I^{\mu_I},
\label{eq:IM}
\end{equation}
where any index value between $1$ and $I_{\max}$ that does not occur in the structure has multiplicity $\mu_I = 0$ and therefore does not appear explicitly in the monomial. For example, in the structure $S$ encoded by Eq.~(\ref{eq:square}), $SQ$ is the only block type with index $I = 3$, void $*$ is the only  type of vertices with index $I = 2$,  and all $7$ other blocks and vertices  have indices $I=1$:  $SQ_2$, $SQ_2'$, $SQ_4$, $1,2,1^*, 2^*$. This means that $\mu_2=\mu_3=1$, and $\mu_1 = 7$,  resulting in the following  IM of the structure $S$:
\begin{equation}
\mathcal{M}(S)=3^1\,2^1\,1^7
\end{equation}

Starting from the IM in Eq.~(\ref{eq:IM}), the B$\&$B complexity can be computed by replacing the summation over object types $\alpha$ in Eq.~(\ref{eq:C}) with a summation over indices:
\begin{align}
&N = \sum_\alpha I_\alpha = \sum_I \mu_I I \\
&C_{B\&B}
= \min_{\text{valid codes}} \left[-\sum_I \mu_I I \log_2\left(\frac{I}{N}\right) \right]
\end{align}

 For each vertex or block type $\alpha$, one may define its “weight’’ in a given SC as 
$w_\alpha = I_\alpha / N$. Since $\sum_\alpha w_\alpha = 1$, this yields
the classical upper bound on entropy:
\begin{align}
\label{eq:ineq}
C_{B\&B}
= -N \sum_{\alpha=1}^{\Omega}
w_\alpha \log_2 w_\alpha \le
N \log_2 \Omega .
\end{align}
Motivated by this bound, we introduce a new measure of block/vertex
diversity,  the \textit{Compositional Complexity}, that only depends on the number of block and bond types,  $\Omega$, and their cumulative usage in the code,  $N$:
\begin{equation}
C_{\text{comp}}
=  N \log_2 \Omega .
\end{equation}

\section{Structural Complexity}

\begin{table}[h!]
\centering
\begin{tabular}{lccccccc}
\hline\hline
Structure & IM
& $\Omega$ & $N$
& $C_{B\&B}$ 
& $C_{\text{comp}}$ & $C_{\text{struct}}$ 
\\
\hline
$\mathrm{C}_{60}$   
  & $1^{4}$ 
  & $4$  & $4$  
  & $8.00$ 
  & $8.00$
  &  $75.75$ 
\\[2pt]
$S$                 
  & $1^{3}\,2^{2}\,3^{1}$ 
  & $6$  & $10$ 
  & $24.46$ 
  & $25.85$
  & $123.77$
\\[2pt]
$(\mathrm{CH}_2\mathrm{OH})_2$ 
  & $1^{4}\,2^{3}$ 
  & $7$ & $10$ 
  & $27.22$ 
  & $28.07$ 
  & $130.41$
\\[2pt]
Zinkblende          
  & $1^{5}\,2^{5}\,4^{1}$ 
  & $11$ & $19$ 
  & $62.71$ 
  & $65.73$
  & $192.72$
\\[2pt]
$\mathrm{Glc}_\alpha$ 
  & $1^{5}\,2^{8}$ 
  & $13$ & $21$ 
  & $76.24$
  & $77.71$ 
  & $224.83$
\\[2pt]
Laves phase         
  & $1^{6}\,2^{18}\,8^{1}$ 
  & $25$ & $50$ 
  & $222.19$ 
  & $232.19$
  & $492.46$
\\
\hline\hline
\end{tabular}
\caption{
Index monomials, number of block and bond types $\Omega$, 
cumulative index $N$, B$\&$B complexity $C_{B\&B}$,  
compositional complexity $C_{\text{comp}}$, and token-based structural complexity $C_{\text{struct}}$ computed for six SCs, Eqs.(\ref{eq:square})- (\ref{eq:LF}).
}
\label{tbl1}
\end{table}

\end{rev_par}
In Table~\ref{tbl1} we list  B$\&$B and 
Compositional Complexities for all the Scs presented above, Eqs (\ref{eq:square})–(\ref{eq:LF}):   the “square’’  $S$, 
 Fig.~\ref{fig:Fig1}c, 
Ethylene Glycol $EG$, Fig.~\ref{fig:Fig1}d), 
the hierarchically encoded $\alpha$-D-glucose, Fig.~\ref{fig:Fig2}c), C$_60$,  
as well as the Zinkblende and Laves phase lattices from 
Ref.~\cite{Jason2025}. 
As expected,  $C_{B\&B}\le C_{\text{comp}} $ for  all cases.  
In fact, this inequality is very tight: the  gap between $C_{B\&B}$ and $C_{\text{comp}}$ is below $5\%$ for all 
examples. This further justifies the use of the simpler quantity, $C_{\text{comp}}$ as a reliable, quick estimate of the compositional complexity. Furthermore, the resulting numerical values correlate well with the apparent length of each Sc.

\begin{rev_par}

\begin{figure}
    \centering
    \includegraphics[width=\linewidth]{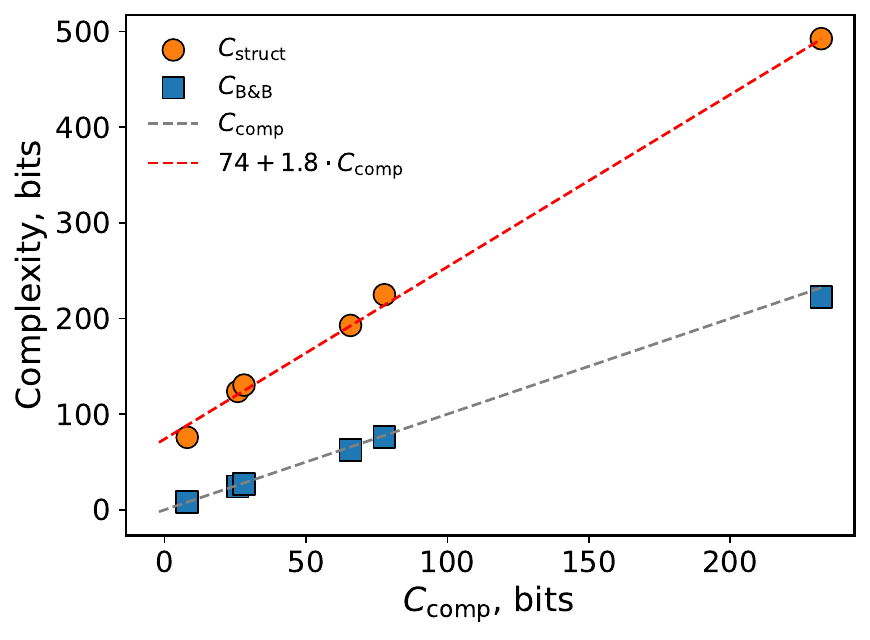}
    \caption{\rev{
    Comparison of three measures of structural information for all structures 
    discussed in the text.  
    Structural complexity  $C_{\text{struct}}$ obtained by direct 
    tokenization of each SC (circles), and 
    $C_{B\&B}$ (squares), plotted 
    against compositional complexity $C_{\text{comp}}$.  
    As expected, 
    $C_{B\&B} \le C_{\text{comp}}$ 
    (blue dashed line).  
    The orange dashed line shows a linear regression of 
    $C_{\text{struct}}$ versus $C_{\text{comp}}$.  
    All three measures exhibit very strong mutual correlations. }
    }
    \label{fig:Fig4}
\end{figure}

Our definition of $B\&B$ Complexity, Eq.~(\ref{eq:C}), implicitly relies on a 
partial “tokenization’’ of the Sc: the process of decomposing the 
SC into elementary informational units (tokens) and assigning an information 
content to each. In practice, we treat \emph{block types} and 
\emph{vertex types} as the primary linguistic units of SC, since they 
correspond directly to physically meaningful ingredients of the structure.  
For simplicity, several syntactic components are omitted from 
$C_{B\&B}$:
(i) declarations of fundamental block types 
(except when introducing a new symmetry variant, as in $B = A.\mathrm{Sym}$);  
(ii) declarations of block symmetries, written “$. \mathrm{Sym}$’’;  
(iii) multiplicity superscripts that appear within composite-block definitions.  
These elements encode additional information, but including them explicitly 
would introduce a level of granularity unjustified by the structural role they 
play and would obscure the hierarchical nature of SC without adding conceptual clarity.

In order to compute the full Kolmogorov Structural Complexity, we need to perform a more comprehensive assessment of the SC information content.  This  
can be achieved by a \emph{full} tokenization of the SC text and 
assigning to each token an information cost based on its frequency.  
The SC format naturally admits such a procedure.  
We replace all original delimiters with semantic prefixes: 
$|$ preceding each block name, 
$+$ and $-$ preceding vertex types $v$ and $v^*$, respectively.  
All remaining delimiters are replaced by “.’’, and every token beginning with “.’’ 
(e.g.\ symmetry groups, multiplicities, orbit sizes) is added to a 
\emph{global} dictionary together with the prefix tokens 
$\{|,+,-\}$.  
This establishes a Kolmogorov-style encoding in which  
block and vertex identifiers use \emph{local} probabilities 
(computed within the given SC), while all other tokens use \emph{global} frequencies.

The Structural Complexity is computed as the total information content of a 
minimal SC that unambiguously defines the structure:
\begin{equation}
C_{\text{struct}} = - \sum_i \log_2 p_i ,
\end{equation}
where the sum runs over all tokens in the Sc and $p_i$ denotes 
either a local or a global token probability, depending on the token type.  
Details of the tokenization rules and probability assignments are provided in 
Appendix~\ref{app:tokenization}.

Formally, $C_{\text{struct}}$ is not a theoretical lower bound but a directly 
computable Shannon information: it is the total information contained in the 
Sc obtained through full tokenization.  As such, it provides an 
estimate of the underlying Kolmogorov complexity of the structure.  In 
principle, a minimal SC must satisfy two conditions: (i) it must uniquely 
specify a single structure, and (ii) no strictly shorter code may do so.  
Establishing these conditions rigorously is generally difficult.  For practical 
purposes, therefore, we characterize the complexity of a structure by the 
information content of the best available Sc, which provides a 
well-defined and reproducible proxy for its algorithmic complexity.

By its definition, $C_{\text{struct}}$ contains $C_{B\&B}$ as the contribution from the local tokens: blocks, and bonds. We can therefore formulate general Complexity Inequalities valid for an arbitrary SC:
\begin{equation}
    C_{B\&B}\lesssim C_{\text{comp}}<C_{\text{struct}}
\end{equation}
As we have seen for all the structures listed in Table \ref{tbl1}, the first of the two inequalities is very tight: $C_{B\&B}$ are $C_{\text{comp}}$ are nearly equivalent for most practical purposes. The gap between $C_{\text{comp}}$ and $ C_{\text{struct}}$ is more significant. This \textit{Complexity Gap} quantifies the amount of additional structural information that is not directly reducible to listing of the blocks and bonds. This information includes the multiplicities of the blocks and their symmetries, including the symmetry of the overall structure.

The calculated values of Structural Complexity $C_{\text{struct}}$ are included in Table \ref{tbl1}, alongside   $C_{B\&B}$,  $C_{\text{comp}}$.  
Figure~\ref{fig:Fig4} compares the three Complexity measures.   
As noted already, $C_{B\&B}$ and $C_{\text{comp}}$ track one another closely.  
Even more striking is the strong correlation between both of these measures and the independently computed token-based Structural Complexity:
\begin{equation}
C_{\text{struct}}
   \;\approx\; 74 \;+\; 1.8\, C_{\text{comp}}
   \;=\;
   74 \;+\; 1.8\, N\log_2 \Omega .
\end{equation}
  
The strong correlation between Structural and Compositional Complexity can
be understood heuristically. In most SCs, the majority of tokens correspond 
either to block names or vertex names, whose frequencies reflect the 
underlying multiplicities of block and bond usage. As a result, the Shannon 
information of the full token sequence is dominated by these frequencies, 
making the combinatorial quantity $C_{\text{comp}}$ an excellent proxy for the 
full information content. 

\end{rev_par}

\section{Conclusions}
In this work, we introduced the \textit{Blocks \& Bonds (B\&B)} hypergraph 
formalism as a unified framework for encoding hierarchical self-assembled 
structures. This approach generalizes classical chemical graphs by capturing 
multi-type interactions, internal symmetries, and recursive modularity.  
Within this framework, the \textit{Structure Code} (SC) provides a compact, 
hierarchical script for specifying assemblies across molecular, colloidal, 
and supramolecular systems.

\begin{rev_par} 
We defined three related measures of structural information.  
The \textit{Structural Complexity} quantifies the minimal information required 
to encode a fully specified assembly, following a Kolmogorov-style logic.  
The closely connected B\&B and \textit{Compositional Complexities} capture the diversity and usage of 
building blocks in the construction set, and are far simpler to compute.

Beyond the development of SC, the central finding of this work is a strong 
correlation among all three complexity measures.  
Across all systems we examined—ranging from small molecules to DNA origami 
and complex crystalline architectures—these quantities track one another with 
remarkable accuracy. 
This observation elevates the simplest of the three measures, the 
Compositional Complexity, to the status of a highly practical tool for 
evaluating the information content of a structure: 
\[
C_{\text{comp}} = N \log_2 \Omega .
\]

The compositional complexity represents the information required to specify the
assignments of $N$ non-equivalent instances drawn from
$\Omega$ distinct block or bond types. The same Shannon-style form appears as a complexity measure across multiple scientific
domains. For instance, in the context of \emph{biocomplexity}, Refs. \cite{ZenilNetworkBiologyReview,Acosta2025BioSignalsReview}
show that maximal sequence, network, or state heterogeneity
scales proportionally to $N\log_{2}\Omega$, where $N$ is the number
of degrees of freedom and $\Omega$ the number of admissible
states. In \emph{social and economic systems}, entropy-based
treatments of diversity
(e.g., in Ref.  \cite{Teza2021})
likewise identify Shannon-type measures as an adequate proxies for 
structural complexity. Finally, in the context of \emph{information science}  \cite{Lloyd2001MeasuresOfComplexity,
FeldmanCrutchfield1998}),  the information
needed to specify a system of $N$ variables over an $\Omega$-symbol
alphabet is precisely $N\log_{2}\Omega$ bits. Within our hypergraph approach, this
interpretation is confirmed empirically for complex hierarchical architectures:  $C_{\mathrm{comp}}$ strongly correlates with the token-base information content.

The B\&B hypergraph representation captures global symmetry, modular 
organization, and stereochemical constraints, while enabling substantial 
compression of structural information.  
This formalism offers a scalable and interpretable language for describing 
hierarchical assemblies and provides a foundation for data-driven 
classification, inverse design, and complexity-aware materials discovery.

\end{rev_par}

\begin{acknowledgments}
 This research was done at and used resources of the Center for Functional Nanomaterials, which is a U.S. DOE Office of Science User Facility, at Brookhaven National Laboratory under Contract No. DE-SC0012704.
\end{acknowledgments}

\bibliography{main}
\newpage

\appendix

\section{Group Theory Primer for Block Symmetries}
The description of blocks in the \textit{Blocks \& Bonds (B\&B)} framework relies heavily on elementary group theory. In this Appendix we provide a short primer, focusing on permutation groups and their role in defining equivalence among block vertices.

\subsection{The Symmetric Group}
The most general group acting on $k$ labeled vertices is the \textit{symmetric group} $S_k$, consisting of all $k!$ possible permutations of the labels $\{1,2,\dots,k\}$. Each element $\sigma \in S_k$ is a bijection from the set of vertices to itself, and group composition is given by permutation composition.
It is convenient to represent elements of $S_k$ groups in cycle notations, e.g. $(123)(45)\in S_6$ corresponds to a group element that permutes cyclically the triad $(1,2,3)$  and the pair $(4,5)$, leaving $6$ untouched.  

As an example, $S_3$ contains $3!=6$ permutations: the identity $e$, the three transpositions $(12),(13),(23)$, and the two 3-cycles $(123),(132)$.

\subsection{Equivalence Groups of Blocks}
A block $B=[k/G]$ is defined by its valence $k$ and an \textit{equivalence group} $G\subseteq S_k$. Two labelings of the block’s $k$ terminals are considered equivalent if they are related by an element of $G$. This captures the notion that certain permutations correspond to symmetries of the block’s geometry or function.

\subsection{Cyclic and Dihedral groups}
In general, $C_k = \langle (1\,2\,3\dots k) \rangle$ is the cyclic subgroup of $S_k$ generated by a single $k$-cycle. The dihedral group $D_k$ combines  $C_k$ with reflection. These groups frequently appear as equivalence groups for planar and polyhedral blocks.

\subsection{The Octahedral Block}
The octahedron has $k=6$ vertices, and its equivalence group is the \textit{octahedral group} $O \subset S_6$, which is a subgroup of all possible permutations. 
This group contains $24$ elements and can be generated, for example, by:
\begin{align}
  \nonumber  r &= (1245), \quad &\text{rotation by $90^\circ$ about $z$-axis} \\
  \nonumber  s &= (123)(456), \quad &\text{$120^\circ$ rotation about the main diagonal}
\end{align}
Together, $r$ and $s$ generate all $24$ even permutations of the octahedron’s vertices that correspond to proper rotations. Thus, the octahedral block is denoted $[6/O]$. 

Within $O$, there exist multiple distinct twofold rotations,  $C_2$. For instance, a $180^\circ$ rotation about the $z$-axis is $(14)(25)$, while a $180^\circ$ rotation about the $xy$-diagonal  $(12)(36)(45)$. 

\subsection{The Square Block}
For a square block $[4/C_4]$, has $k=4$ vertices, with equivalence group $C_4$. This implies that there are four equivalent resresntations of the same block, different by rotation: $(1,2,3,4)\sim(2,3,4,1)\sim(3,4,1,2)\sim(4,1,2,3)$. Since the equivalence group does not include reflection, $(1234)$ is not equivalenmt to $1432$. In contrast, a completely symmetric square block $[4/C_4].D_4$, such as (1,1,1,1) posseses full $D_4$ symmetry.   
Analogous to the octahedron, there are multiple inequivalent $C_2$ subgroups of $D_4$:  
$(13)(24)$ corresponds to a $180^\circ$ rotation about the square’s center;  
$(12)(34)$ corresponds to reflection symmetry across a vertical axis; $(14)(23)$ corresponds to reflection across the diagonal.  

\begin{rev_par}

\section{ Specification of the Structure Code}
\label{app:structurecode}

This Appendix provides a complete, minimal specification of the
\emph{Structure Code} used to encode finite or periodic B\&B
(Block--and--Bond) hypergraphs. The code defines:  
(i) vertex types;  
(ii) fundamental and composite blocks;  
(iii) bonds via complementary vertex types;  
(iv) hierarchical block composition;  
(v) optional crystalline periodicity.

\subsection{Primitive Objects}

\paragraph{Vertex types.}
Each vertex is assigned a type from the set  
\(\mathcal{V}=\{0,1,2,\dots\}\cup\{0^*,1^*,2^*,\dots\}\cup\{*\}\).  
Types \(v\) and \(v^*\) are complementary and denote the endpoints of a bond.  
The special type \(*\) is a non-interacting terminal.

\paragraph{Blocks.}
A block \(B\) of valence \(k\) is a hyperedge with \(k\) terminals.  
Its label has the form
\[
  B = [k/G].\mathrm{Sym},
\]
where \(G\subseteq S_k\) is an equivalence group acting on the terminals  
and \(\mathrm{Sym}\subseteq G\) is the symmetry group of the block instance.  
If the terminals split into orbits of sizes \((k_1,\dots,k_\ell)\) under \(G\),
we write \([ (k_1,\dots,k_\ell) / G ]\).

A block is \emph{fundamental} if all terminals are bare vertex types.  
It is \emph{composite} if it is constructed from other blocks.

\paragraph{Bonds.}
Bonds are implicit: each vertex of type \(v\) must match exactly one vertex of
type \(v^*\).  If $v^*$ is not present in the code,  type $v$ matches its own type. By default, no other matching is allowed.

\subsection{SC Syntax}

A SC consists of:
(i) block-type declarations;  
(ii) a top-level block or crystal definition.

\paragraph{Fundamental block declaration.}
\[
  B = [k/G].\mathrm{Sym}
\]
where \(\mathrm{Sym}\) is optional.

\paragraph{Composite block declaration.}
Composite blocks use the Block–Vertex–Multiplicity–Symmetry (BVMS) form:
\[
  B = [ T_1 \;|\; T_2 \;|\; \dots ]. \mathrm{Sym},
\]
where each term \(T\) has the form
\[
  T = B_i(v_1,\dots,v_n)^{m},
\]
with:
\begin{itemize}
  \item \(B_i\): a previously defined block type (fundamental or composite);
  \item \((v_1,\dots,v_n)\): vertex types assigned to the terminals of \(B_i\);
  \item \(m\): multiplicity (default \(m=1\));
  \item terms separated by ``\(|\)'' correspond to contributions from distinct block types.
\end{itemize}

\paragraph{Hierarchy.}
Any composite block may be used as a building unit when defining a higher-level block. Definitions must not be recursive.

\subsection{Crystalline Structures}

Periodic structures are encoded per unit cell:
\[
  B = [ T_1 \;|\; T_2 \;|\; \dots ]^{\infty}.\mathrm{Sym},
\]
where the subscript \(\infty\) denotes infinite repetition, and
\(\mathrm{Sym}\) is a crystallographic space group in Hermann--Mauguin notation.  
All vertex types with complements must match within the periodic arrangement.

\subsection{Reserved Symbols and Conventions}

\begin{itemize}
  \item Block names are alphanumeric strings (e.g.\ \verb|C|, \verb|OCT|, \verb|B1|).
  \item Vertex multiplicity is denoted by a superscript:
        \((v_1,\dots,v_n)^m\).
  \item Groups \(C_n,D_n,A_n,S_n,O,O_h\), etc.\ follow standard mathematical conventions.  
  \item ``\(|\)'' separates contributions of different block types within a composite definition.  
  \item ``\(*\)'' is a neutral terminal and never participates in bonding.
  \item Complementary vertex types must pair one-to-one globally.
\end{itemize}

\section{Appendix: Tokenization Procedure for Structure Codes}
\label{app:tokenization}

This Appendix describes the tokenization procedure used to estimate the
information content of each Structure Code (SC).  The goal is to construct,
for each structure, a linear sequence of tokens
\(\{t_1, t_2, \dots, t_{N_t}\}\), and then assign to each token an information
content based on either local or global occurrence probabilities.

\subsection{Token Types and Classes}

We distinguish two classes of tokens:

\begin{itemize}
    \item \textbf{Local tokens}:
    names of block types and vertex types. Their probabilities are computed
    separately for each SC.
    \item \textbf{Global tokens}:
    the three prefix symbols
    \(\{ +, -, | \}\), and all tokens of the form ``.X’’ 
    (i.e.\ those beginning with a dot). These include group labels,
    exponents, orbit sizes, space groups, etc. Their probabilities
    are computed once from the combined ensemble of all seven SCs.
\end{itemize}

The dot character ``.'' itself does not appear as an isolated token: it is
always absorbed into the following symbol, as explained below.

The void terminal ``*'' is treated specially: it produces only a prefix
token ``+’’ with no associated local symbol.

\subsection{Construction of the Token Sequence}

Starting from the plain-text Structure Code, we apply the following steps:

\begin{enumerate}
    \item \textbf{Base splitting.}
    The code string is split into fragments using the structural delimiters
    \([\, ], (\, ), \{\,\}, =, |, ,, ;, ., / \) as separators.
    Empty fragments are discarded.

    \item \textbf{Prefix assignment.}
    Each fragment is classified and converted into one or two tokens:
    \begin{itemize}
        \item If the fragment is a block name (e.g. A, B, C, Oct), it yields
        a \emph{block prefix} token ``|’’ followed by a \emph{local}
        block token.
        \item If the fragment is a vertex body (an integer $v$,
        optionally followed by ``*'', or the bare ``*''):
        \begin{itemize}
            \item bare integer $v$ $\to$ prefix ``+’’ and local vertex token $v$;
            \item $v*$ $\to$ prefix ``-’’ and local vertex token $v$;
            \item bare ``*’’ $\to$ prefix ``+’’ only (no local token).
        \end{itemize}
        \item All other fragments (group names, orbit sizes, exponents,
        space-group symbols, etc.) are turned into \emph{global} tokens by
        prefixing them with a dot, i.e.\ ``.X’’.
    \end{itemize}

    Thus, at this stage the token stream consists of the three prefix symbols
    \(|, +, -\) and a collection of block names, vertex labels, and raw
    fragments.

    \item \textbf{Dot absorption.}
    Any occurrence of the dot symbol ``.'' in the original text is absorbed
    into the following token: instead of a separate token ``.'' followed
    by \text{X}, we create a single global token ``.X’’. In the final
    token streams, there is no standalone ``.'' token; all dot-prefixed
    tokens are of the form ``.X'' and treated as global.
\end{enumerate}

After these steps, each SC is represented as a sequence whose elements are:
\begin{itemize}
    \item global prefix tokens: \(|, +, -\);
    \item local block tokens: block names, always immediately following ``|'';
    \item local vertex tokens: integer labels, always immediately following
          ``+’’ or ``-’’;
    \item other global tokens: all $.X$ strings:
          $.S_3$,  $.A\_4$, $.I_h$, $.Fd3m$, etc.
\end{itemize}

As an illustration, we show the token sequences for C$_{20}$ and ethylene
glycol (EG), using space-separated tokens in a \text{typewriter} font.
\paragraph{C\texorpdfstring{\(_{20}\)}{} dodecahedron.}
Starting with SC
\[
C = [3/S_3];\quad C20 = [C(1)^{20}].I_h
\]
the token sequence is
\[
| C + 3 .S_3 | C20 | C + 1 .^{20} .I\_h
\]

\paragraph{Ethylene glycol \((\mathrm{CH}_2\mathrm{OH})_2\).}
From SC
\[
\begin{aligned}
& C=[4/A_4].C2;\quad O=[2/C_2];\quad H=[1];\\
& EG=[C(0,2,3)^2\,|\,O(2^*,1)^2\,|\,H(0^*)^4,(1^*)^2].C_2
\end{aligned}
\]
We obtain the token sequence
\begin{align}
&| C + 4 .A_4 | C2 | O + 2 .C_2 | H + 1 \\
\nonumber &| EG |C + 0 + 2 + 3 .^2 | O - 2 + 1 .^2 | H - 0 .^4 - 1 .^2 .C_2
\end{align}

Tokenized versions for the remaining structures (C$_{60}$, $S$, Glc$_\alpha$,
Zinkblende, and the Laves phase) are obtained in exactly the same way and were
used in the numerical analysis below.

\subsection{Information Content of a Structure Code}

For each SC we build a local dictionary of block and vertex tokens and
compute their local probabilities
\[
p^{(\mathrm{loc})}(x) =
\frac{n^{(\mathrm{loc})}(x)}{N^{(\mathrm{loc})}},
\]
where \(n^{(\mathrm{loc})}(x)\) is the number of occurrences of token $x$
among block and vertex tokens in that SC, and
\(N^{(\mathrm{loc})}\) is the total number of local tokens in that SC.

In parallel, a global dictionary is built by aggregating all occurrences of
prefix tokens \(|,+,-\) and all dot-prefixed tokens ``.X'' across the seven
SCs. Their global probabilities are
\[
p^{(\mathrm{glob})}(y) =
\frac{n^{(\mathrm{glob})}(y)}{N^{(\mathrm{glob})}},
\]
where \(n^{(\mathrm{glob})}(y)\) counts occurrences of the global token $y$
in the combined corpus, and \(N^{(\mathrm{glob})}\) is the global total.

For each token \(t_i\) in the sequence of a given SC, the probability
assigned in the information calculation is
\[
p_i =
\begin{cases}
p^{(\mathrm{loc})}(t_i), & \text{if $t_i$ is a block or vertex token},\\[4pt]
p^{(\mathrm{glob})}(t_i), & \text{if $t_i$ is a prefix or dot-prefixed token}.
\end{cases}
\]
If an SC contains \(N_t\) tokens in total, its information content is
defined as
\begin{equation}
C_{\text{struct}}
=
-\sum_{i=1}^{N_t} \log_2 p_i,
\label{eq:I_info}
\end{equation}
where the probabilities \(p_i\) are assigned according to the scheme above.
This mixed local/global assignment reflects the idea that block and vertex
labels are structure-specific (local), while the syntactic symbols and
group-related tokens form a global “language’’ shared across all codes.

Table~\ref{tbl:info} summarizes, for each of the seven structures considered
in the main text, the total number of tokens \(N_t\) in its SC and the
resulting information content \(C_{\text{struct}}\) computed via
Eq.~(\ref{eq:I_info}).

\begin{table}[h!]
\centering
\begin{tabular}{lcccc}
\hline\hline
Structure & $N_t$  & $N_{\mathrm{loc}}$ & $C_{\text{struct}}$ (bits) & $C_{\text{struct}}/N_t$ \\
\hline
C$_{60}$          & 25  & 9  & 75.75 & 3.03 \\
$S$               & 41  & 16 & 123.77 & 3.02 \\
EG                & 43  & 18 & 130.41 & 3.03 \\
Zinkblende        & 62  & 27 & 192.72 & 3.11 \\
Glc$_\alpha$      & 80  & 30 & 224.83 & 2.81 \\
Laves phase       & 134 & 57 & 492.46 & 3.68 \\
\hline\hline
\end{tabular}
\caption{Total number of tokens $N_t$, number of local tokens $N_{\mathrm{loc}}$
(block and vertex labels), and information content $C_{\text{struct}}$ for each
Structure Code, computed using the mixed local/global token probabilities
described in Appendix~\ref{app:tokenization}. The last column shows the
average information per token.}
\label{tbl:info}
\end{table}

\end{rev_par}

\end{document}